\documentclass[manuscript,screen,nonacm]{acmart}
\usepackage[ruled,linesnumbered]{algorithm2e}
\usepackage{amsmath, amsthm}

\SetAlFnt{\small\sf}
\makeatletter
\newcommand{\nosemic}{\renewcommand{\@endalgocfline}{\relax}}
\newcommand{\dosemic}{\renewcommand{\@endalgocfline}{\algocf@endline}}
\newcommand{\pushline}{\Indp}
\newcommand{\popline}{\Indm\dosemic}
\let\oldnl\nl
\newcommand{\nonl}{\renewcommand{\nl}{\let\nl\oldnl}}
\makeatother
\AtBeginDocument{%
  \providecommand\BibTeX{{%
    \normalfont B\kern-0.5em{\scshape i\kern-0.25em b}\kern-0.8em\TeX}}}

\setcopyright{none}

\begin{document}

\title{Explicit caching HYB: a new high-performance SpMV framework on GPGPU  }

\author{Chong Chen}
\authornote{correspond author.}
\affiliation{%
  \institution{National Institute of Supercomputing, University of Nevada, Las Vegas}
  \streetaddress{4505 S Maryland Parkway}
  \city{Las Vegas}
  \state{Nevada}
  \country{USA}
  \postcode{89119}
}
\email{chong.chen@unlv.edu}

\renewcommand{\shortauthors}{Trovato and Tobin, et al.}

\begin{abstract}
Sparse Matrix-Vector Multiplication (SpMV) is a critical operation for the iterative solver of Finite Element Methods on computer simulation. Since the SpMV operation is a memory-bound algorithm, the efficiency of data movements heavily influenced the performance of the SpMV on GPU. In recent years, many research is conducted in accelerating the performance of SpMV on the graphic processing units (GPU). The performance optimization methods used in existing studies focus on the following areas: improve the load balancing between GPU processors, and reduce the execution divergence between GPU threads. Although some studies have made preliminary optimization on the input vector fetching, the effect of explicitly caching the input vector on GPU base SpMV has not been studied in depth yet.In this study, we are trying to minimize the data movements cost for GPU based SpMV using a new framework named” explicit caching Hybrid (EHYB)”. The EHYB framework achieved significant performance improvement by using the following methods: \par
1. Improve the speed of data movements by partitioning and explicitly caching the input vector to the shared memory of the CUDA kernel. \par
2. Reduce the volume of data movements by storing the major part of the column index with a compact format. \par
We tested our implementation with sparse matrices derived from FEM applications in different areas. The experiment results show that our implementation can overperform the state-of-the-arts implementation with significant speedup, and leads to higher FLOPs than the theoryperformance up-boundary of the existing GPU-based SpMV implementations.
\end{abstract}



\maketitle
\pagestyle{plain}
\section{Introduction}
Sparse Matrix-Vector Multiplication (SpMV) evaluates the results of \(y = A\times x\), where \(A\) is a sparse matrix and \(x\) and \(y\) are (usually) dense 
vectors. SpMV is a critical operation for a wide range of scientific computing applications, especially in the iterative solver for the very large sparse 
linear system derived from the partial differential equation (PDE) in Finite Element Methods (FEM). 
With the computing capability for single-core processors limited by the 
power consumption problem that came with increasing frequency scaling \cite{mittal2015survey}, using the 
massive parallel computing processor such as general-purpose graphic 
processing units (GPGPU) became a must to improve the performance of 
scientific computing applications. For the important role of SpMV in scientific
computing, the efficient parallel implementation of SpMV on the 
GPGPU architectures has been a hot research topic in the field of high-performance computing\cite{anzt2020load}. And some of the new developed algorithms was
integrated into the CUSPARSE library provided by NVIDIA.\par 
In this study, we are work on developing a  
GPGPU based SpMV implementation leads to better performance than the
state-of-art implementation of SpMV on GPU with reasonable 
preprocessing time. We try to optimizing the SpMV using 
the following methods:\par
\begin{itemize}
  \item Reduce the amount of data movements by reduce the 
  I/O volume for SpMV and 
  \item Approaching the high efficiency of data movements by 
  partitioning and explicitly caching the 
input vector to the shared-memory of GPU device.
\end{itemize}
The above methods are accomplished through a "partitioning, reordering, 
and caching" procedure, categorized as preprocessing of 
the sparse matrix. The time cost of preprocessing procedure is 
comparable to the time cost for the format change of the sparse
matrix, and the partitioning/reordering will introduce an extra 
preprocessing time. And the preprocessing time costs, based on or experiment, around 400$\times$ to 2000$\times$ of a single SpMV on GPU. 
\par
We tested our implementation with large sparse matrices derived from FEM  
on structure, biomedical, Computational Fluid Dynamic (CFD), and electromagnetics problems.
Most of these matrices are generated with an unstructured mesh (which leads to an
irregular sparse pattern). Based on our experiment on the  
Tesla V100 GPU, the optimized SpMV kernel we developed 
can leads to better performance when compared to the state-of-the-art GPU based library functions on SpMV.  \section{Background}
\label{section2}
\subsection{Concepts related to GPU code optimization}
Implementing SpMV in GPGPU efficiently requires the code to be optimized 
according to the GPGPU architecture. Since the CUDA programming model 
becomes the de facto for GPGPU parallel programming, all the 
architecture-related code optimization discussed in this study will be based on the 
NVIDIA GPGPU and CUDA program model. \par
In the CUDA programming model, the GPU code is offloaded to the devices through
\emph{kernel function} and all the kernel functions executed the instructions 
which will be executed with a \emph{block} of threads. The block will
be divided into \emph{warps}, which is a group of 32 threads. The GPU 
\emph{stream multiprocessor (SM)} will try to execute the instructions belongs 
to one warp at each clock cycle. It is common that multiple warps of 
the same block were dispatched to one SM, and the pipeline of that SM will switch
between the contexts of warps when the current warp is stall (i.e. waiting for data). 
Threads within the same warp execute different instructions will
cause \emph{thread divergence}. When thread divergence happened, the instructions 
from multiple threads will be executed serially.\par  
GPU memory hierarchy is different compared to CPU memory hierarchy. 
According to the terminologies of CUDA, GPU memory space can be briefly 
categorized as: shared memory L1/texture cache, register file, 
L2 cache and global memory. Register files are the memory space used for 
stack variables of the CUDA threads. Global memory usually represents 
the RAM which is installed outside the GPU chip. The global memory is connected
to the GPU chip wich DDR/HBM memory interface with limited bandwidth
(which is 900 GB/s for Tesla V100 used in this study) and the latency of 
global memory can be hundreds of clock cycles of GPU chip. 
The L1/texture cache and shared memories are "fast memory" which is consider 
as a part of the SM, these memory are connected to the SM processing units with a 
verh high bandwidth connections.  
 \par
maximize the hit rate of the input vector fetch of SpMV operations. 
\subsection{Related research about SpMV on GPU}

Significant research about optimizing the SpMV performance on GPGPU is conducted because SpMV is important in a large range of computing applications.
This research starts at \cite{bell2009implementing}, where the researchers start to propose different sparse matrix storage formats to reduce the 
imbalance of GPU programming on matrices with different nonzero patterns. The format for high-performance SpMV on GPU examined by \cite{bell2009implementing} 
includes Ellpack(ELL), Compressed Row(CSR), Coordinate(COO),  Diagonal(DIA), and Hybrid(HYB). Numerous of research is conducting on finding the best format 
automatically for different blocks, and optimal partitioning the matrix according to the nonzero patterns using statistical 
methods\cite{li2014performance}\cite{yang2017hybrid}\cite{yang2018parallel}, or machine learning 
methods\cite{dufrechou2021selecting}\cite{nisa2018effective}\cite{benatia2016sparse}. There are also research focus on 
developing new formats which may leads to better 
balancing on a wide range of matrices, the proposed novel format includes SELLp\cite{anzt2014implementing}, BiELL\cite{zheng2014biell}, 
and JAD\cite{cevahir2009fast}, to name a few. Among these novel formats, the best performance format is the BCOO format proposed 
by the yaspmv framework\cite{yan2014yaspmv}, however, the BCOO format requires an extremely long time of preprocessing (averagely 155,000$\times$)\par 
In recent year, new algorithms are developed for efficiently conducting GPU based SpMV for imbalanced matrices with CSR format. 
Such as CSR5\cite{liu2015csr5}, merge-based SpMV\cite{merrill2016merge}, and hola SpMV\cite{steinberger2017globally}. These new algorithms can 
provides excellent performance in a variety of nonzero patterns without format conversion. In the latest NVIDIA CUSPARSE library, 
a generic interface for SpMV is introduced. According to \cite{anzt2020evaluating}, the generic interface SpMV kernel in CUSPARSE can overperform many
GPU libraries without conversion.\par
However, research on improving the input vector fetch efficiency on GPU-based SpMV is limited.  In this study, we proposed a new SpMV framework on GPU
named Explicitly Caching Hybrid framework (EHYB), multiple optimization methods is applied to make this framework overperform the state-of-the-art
SpMV functions on GPUs. 
\section{The framework of EHYB on GPU}
\begin{figure*}[ht]
  \centering
  \includegraphics[width=\linewidth]{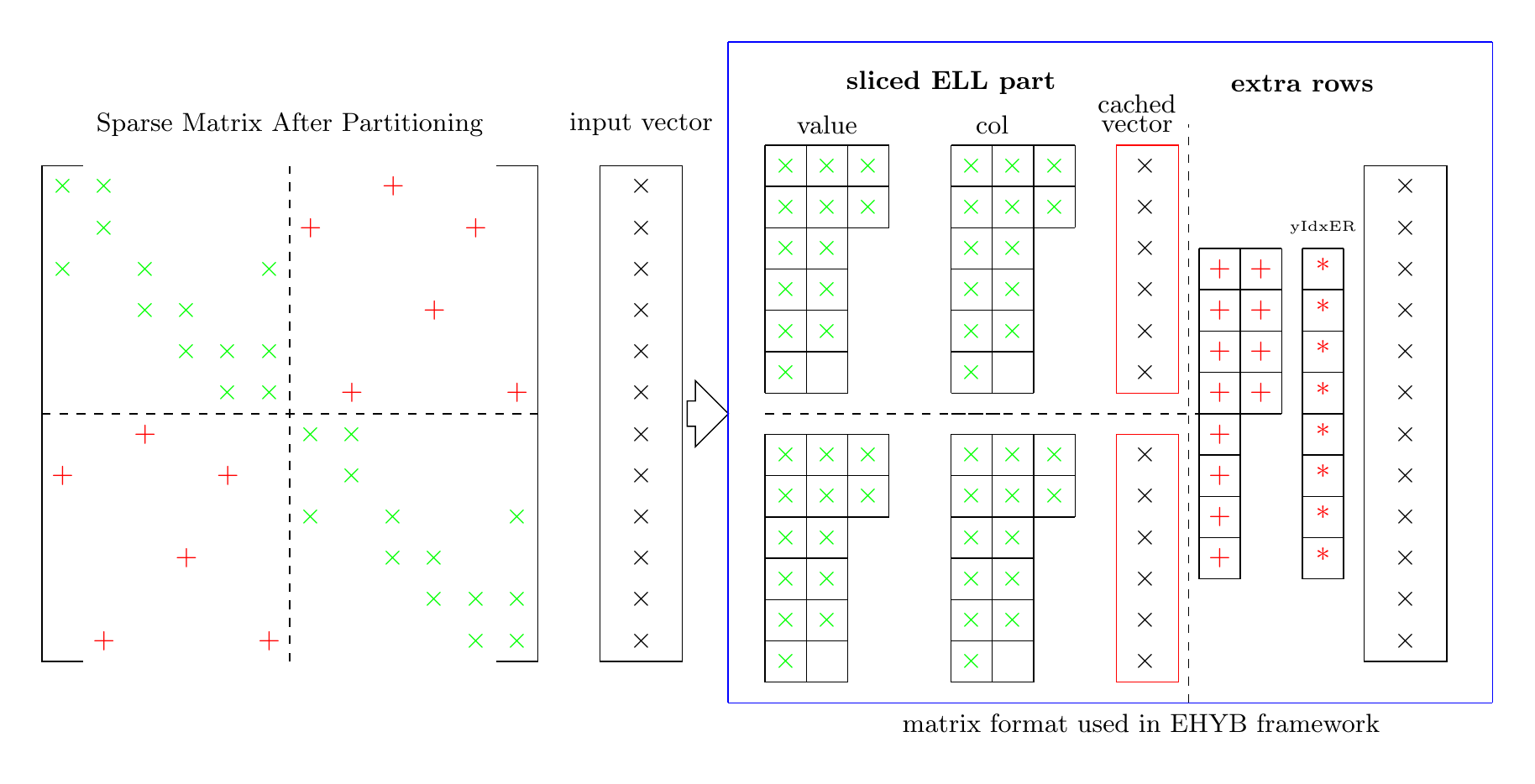}
  \caption{ A sparse matrix in EHYB framework}
  \label{fig:EHYB format}
  \Description{Format description of EHYB}
\end{figure*}
\subsection{Graph-based partitioning as the first step of preprocessing }
Figure \ref{fig:EHYB format} demonstrated the basic concepts of the EHYB framework. The first phase of the EHYB framework is to apply the graph-based 
partitioning to the input sparse matrix. Before partitioning, the sparse matrix will be recognized as an undirected graph with each row/column as a vertex 
and each entry as an edge connected the two vertexes related to its coordinate. After partitioning, the entries of the sparse matrix will be assigned to 
certain partitions, and their row index and column index as vertexes are most likely belong to the same partition. The left part of the figure
\ref{fig:EHYB format} represents a sparse matric after the graph-based partitioning is applied. The SpMV operations on the major part of the 
entries in that matrix will only require input vector data with the index belongs to its partition (which was demonstrated as green \textcolor{green}
{$\times$} in the sparse matrix in figure \ref{fig:EHYB format}). 
However, the partitioning can't be perfect, which means there are still a few entries do not fall within this category, and these entries is demonstrated as 
\textcolor{red}{$+$} in the figure. For these entries, the SpMV operations will require the input vector value with its index out of the partition of that entry.  \par
For the entries represent as green \textcolor{green}{$\times$} in figure \ref{fig:EHYB format}, we can explicitly cache the input vector related to these 
entries in the shared memory of GPU. After the data is explicitly cached, the SpMV operations on the major part of the sparse matrix entries will not 
fetching data from the device memory of GPU. Instead of that, they will directly got the input vector value from the shared memory of GPU with much less latency.\par 
Please be aware that the hypergraph-based partitioning is not suitable for the scenarios in this study. The hypergraph-based partitioning is developed for 
the SpMV operations running on a distributed memory system. For that scenario, the computing node only needs to fetching the same input vector value from 
other nodes once, because it can reuse the fetched data at the local memory. This premise is no longer valid in GPU architecture, which can be categorized as 
a shared memory system with an independent cache for each stream processor. In GPU, the input vector value fetched from other partitions will not stay in the 
cache for a long term. For that reason, the graph-based partitioning algorithm is more suitable for the EHYB framework in this paper. \par
\subsection{The hybrid format of sparse matrix}
To further improve the performance of the CUDA kernel function. The EHYB framework will store the sparse matrices to the format demonstrated as the left 
part of figure \ref{fig:EHYB format}. From this figure we can find out that, the sparse matrix is split into two parts. The sliced ELL part stores
the entries which only require input vector value from the cache, and the \emph{stride} of the slice will be equal to the size of warp in GPU (which is 32). 
The post-partitioning reordering is conducted in a way that the reordered matrix will have its number of entries in row arranged in descending order within certain partitions. This reordering will reduce the thread divergence of CUDA warps since each thread belong to a warp will execute equal number of
iterations on each slice of the matrix. The entries belong to the same partition will be processed by a single CUDA block because they need to share
the input vector stored in the same cache. \par
The \textbf{extra rows (ER)} part of the EHYB matrix in figure \ref{fig:EHYB format} is also stored in an order according to its number of entries in the row. 
However, the input vector in the ER part of the matrix will not be cached into the shared memory. Also, an array of index \emph{yIdxER} should be introduced for the ER part of the matrix, because the re-arrangement of the data is not a \emph{reordering}, so we need an array to map the address of the ER data line to the real row number of the matrix. \par 
The details about the implementation of preprocessing and CUDA kernels of EHYB will be demonstrated in the next section of this paper.
\subsection{Input vector cache size}
The number of partition and size of the input vector cache of EHYB framework can be determined using the following equation:
\begin{equation}
  \label{eq_k}  
  K = MIN_{K\in \mathbb{Z}}(\frac{dimension\times\tau}{K\times P}<SHM^{max})
\end{equation}
\begin{equation}
  \label{eq_v}  
  VecSize = \frac{dimension}{K\times P}
\end{equation}
\par
Where $K$ is an integer value, $dimension$ is the number of rows in the matrix. $\tau$ is the number of bytes per value ($\tau = 4$ for single precision)
and $\tau = 8$ for double precision). $P$ is the number of processor units of the GPU device. $P\times K$ is the number of partition of EHYB, and the input
vector size can be calculated tough equation \ref{eq_v} 
The derivation of the equation \ref{eq_k} and \ref{eq_v} is straightforward, to maximize the performance of the SpMV kernel, we want to utilize all computing
units (which is \emph{streaming processor} in NVIDIA terminology) of the GPU. Also, we want to minimize the size of the ER entries of the EHYB matrix. So 
we set the vector cache size to the largest value which can make the number of partition equal to an integer which is multiple of the number of processing units in the GPU(which 80 \textbf{SMX} in the Tesla V100 GPU used in this study) . 
\subsection{reduce the memory footprint size of matrix}
In figure \ref{fig:EHYB format}, the \emph{col} data of the sliced ELL part of the EHYB format stores the index of the input vector value the CUDA kernel 
will fetch during the SpMV operations. Consider index is used for fetching data from the cached vector in shared memory, and the vector cache size is
limited by equation \ref{eq_k} and \ref{eq_v}, we can confirm that the index value can't be higher than $2^{16}$. This feature gives us the possibility
to continue optimizing the EHYB framework by store the \emph{col} data in figure \ref{fig:EHYB format} in short integer format. This will reduce the 
memory footprint of sliced ELL part by 25\% in single precision, or 13.3\% in double precision. 
\section{The implementation of EHYB}
This section describes the detail of the implementation of preprocessing and CUDA kernel function of the EHYB framework. The ALgorithm \ref{algPrep}
and \ref{alg:2} demonstrates the preprocessing phase of EHYB running on the host machine (CPU) and the CUDA kernel function described in algorithm \ref{alg:3}
will be executed on the GPU.
\subsection{Preprocessing and matrix reordering for the EHYB framework}

\begin{algorithm}[h]
\SetKwData{ReorderTable}{ReorderTable}
\SetKwData{ArrangeTable}{ArrangeTable}
\SetKwData{ColELL}{ColELL}
\SetKwData{ValELL}{ValELL}
\SetKwData{PositionELL}{PositionELL}
\SetKwData{PositionER}{PositionER}
\SetKwData{WidthER}{WidthER}
\SetKwData{PartVec}{PartVec}
\SetKwData{PartId}{PartId}
\SetKwData{row}{row}
\SetKwData{col}{col}
\SetKwData{InputMatrix}{InputMatrix}
\SetKwData{yIdxER}{yIdxER}
\SetKwData{Dimension}{Dimension}

$G(V, E)$ $\leftarrow$ \InputMatrix\;
\SetAlgoLined
\PartVec $\leftarrow$ ParMETIS($G(V, E)$)\;
$Struct$ $S \{int rowIdx; int entries;\}$\;
\For{\row from $1$ to \Dimension}{
  \For{\col from $1$ to $nnz$ at \row }{
    S\_array1[row].rowIdx=\row\;      
    \uIf{\PartVec$[\col]==$ \PartVec$[\row]$}{
      S\_array1[\row].entries$+=1$\;      
      \lIf{S\_array1[row].rowIdx$==0$}{S\_array1[row].rowIdx=\row}      
    }
    \Else{
      S\_array2[].entries$+=1$\;      
      \lIf{S\_array2[row].rowIdx$==0$}{S\_array2[row].rowIdx=\row}      
    }
  }
}
sort(S\_array2) according to entries\;
\For{\PartId from $1$ to patition}{
  sort(S\_array1[\PartId$\times vectorCacheSize$ to (\PartId + 1)$\times vectorCacheSize$])\; 
}
\For{\row from $1$ to \Dimension}{
  \ReorderTable[S\_arry1[\row].rowidx]$\leftarrow$\row\;
}
\For{\row from $1$ to $ER\_rowNumber$}{
  ArrangeTable[S\_array2[\row].rowidx]$\leftarrow$\row\;
  yIdxER[\row]$\leftarrow$ReorderTable[S\_array2[\row].rowidx]
}
Generate the \PositionELL and \PositionER vector similar to SELLp\cite{anzt2014implementing}\;
\caption{Preprocessing algorithm of EHYB framework}
\label{algPrep}
\end{algorithm}
Algorithm \ref{algPrep} of this paper demonstrated the preprocessing phase 1 of the EHYB framework. This algorithm will generate several "metadata" vectors 
for the EHYB framework. The input of this algorithm is a sparse matrix with the coordinate (COO) format. In line 2 of algorithm \ref{algPrep}, the sparse
matrix is converted to a graph and passed to a function of multi-thread METIS library for the graph-based partitioning. The partitioning will generate a 
vector (\emph{PartVec}) which indicates the partition assigned to each vertex of the graph.\par 
Line 3 to line 14 of algorithm \ref{algPrep} created the struct array which stores the number of entries in each row for the sliced ELL part and ER part of the
EHYB matrix. This array will be used as the input of sort operations in line 16 and line 18. Please notice that the for loop in line 17 is the main 
difference between the reordering in EHYB framework and the regular METIS-based reodering. In this for loop, the reorder permutation is determined 
by the descending order of the number of entries in row. Line 20 to line 27 of algorithm \ref{algPrep} will generating the meta vector for the metadata 
vector for the reodering and re-arrangement operations.\par  
After the metadata vectors for the EHYB format matrix are generated, the reordering of the matrix will be completed through Algorithm \ref{alg:2}. In 
line 4 to line 5 of algorithm \ref{alg:2}, the post-partitioning reordering will be conducted as the value and col index of matrix moved to the sliced ELL
part of the HYB matrix. And in line 10 to line 13 of the Algorithm \ref{alg:2}, the re-arrangement will move the entries of the matrix to the ER part of the 
EHYB matrix.
\begin{algorithm}[ht]
  \SetKwData{ReorderTable}{ReorderTable}
  \SetKwData{ArrangeTable}{ArrangeTable}
  \SetKwData{ColELL}{ColELL}
  \SetKwData{ValELL}{ValELL}
  \SetKwData{ColER}{ColER}
  \SetKwData{ValER}{ValER}
  \SetKwData{PositionELL}{PositionELL}
  \SetKwData{WidthELL}{WidthELL}
  \SetKwData{ColELL}{ColELL}
  \SetKwData{ValELL}{ValELL}
  \SetKwData{PositionER}{PositionER}
  \SetKwData{WidthER}{WidthER}
  \SetKwData{PartVec}{PartVec}
  \SetKwData{PartId}{PartId}
  \SetKwData{row}{row}
  \SetKwData{col}{col}
  \SetKwData{InputMatrix}{InputMatrix}
  \SetKwData{EntryAtRow}{EntryAtRow}
  \SetKwData{EntryAtRowER}{EntryAtRowER}
  \SetKwData{Dimension}{Dimension}

  \SetAlgoLined
  \For{$row$ from $1$ to \Dimension}{
    $k1\leftarrow0$;$k2\leftarrow0$;\; 
    \For{in }{
      \uIf{\PartVec$[col]$ }{
        \nosemic$ELL\_idx\leftarrow\PositionELL[\ReorderTable[row]]$\;
        \pushline\dosemic\nonl$+\ReorderTable[row]\%warpSize + k1 \times warpSize$\;
        \popline\ColELL[$ELL\_idx$]$\leftarrow$\ReorderTable$[col]$\;
        \ValELL[$ELL\_idx$]$\leftarrow$\InputMatrix$[row][col]$\;
        $k1+=1$ 
      }
      \Else{
        \nosemic$ER\_idx\leftarrow\PositionER[\ArrangeTable[row]]$\;
        \pushline\dosemic\nonl$+\ArrangeTable[row]\%warpSize + k2 \times warpSize$\;
        \popline\ColER$[ER\_idx]\leftarrow$\ReorderTable$[col]\;$\;
        \ValER$[ELL\_idx]\leftarrow$\InputMatrix$[row][col]$\;
        $k2+=1$ 
      }
    }
  }
  \caption{Reordering phase of EHYB framework}
  \label{alg:2}
  \end{algorithm} 

\subsection{Balancing optimization in CUDA kernel}
The format we utilized in EHYB is similar to SELLp, which may cause unbalancing on the SpMV operations. To overcome this problem, we introduced balancing 
optimization in the CUDA kernel of EHYB, which is demonstrated in algorithm \ref{alg:3}.\par  
In line 6 to line 12 of the algorithm \ref{alg:3} the warp of a CUDA threads will processing the SpMV operations on a \emph{slice} of entries. After the warp
finished processing the current slice, line 14 to line 17 of algorithm \ref{alg:3} will assign a new slice belonging to the same partition to this warp. 
The kernel will processing the ER part of the EHYB matrix with same routine. The only difference is when processing ER part, the kernel will find the 
next slice globally, instead of finding the slice from same patition. 
\begin{algorithm}[]
  \SetKwData{CachedVec}{CachedVec}
  \SetKwData{ColELL}{ColELL}
  \SetKwData{ValELL}{ValELL}
  \SetKwData{ColER}{ColER}
  \SetKwData{ValER}{ValER}
  \SetKwData{PositionELL}{PositionELL}
  \SetKwData{WidthELL}{WidthELL}
  \SetKwData{ColELL}{ColELL}
  \SetKwData{ValELL}{ValELL}
  \SetKwData{PositionER}{PositionER}
  \SetKwData{WidthER}{WidthER}
  \SetKwData{InputVector}{InputVector}
  \SetKwData{OutputVector}{OutputVector}
  \SetKwData{PartId}{PartId}
  \SetKwData{row}{row}
  \SetKwData{InputMatrix}{InputMatrix}
  \SetKwData{Dimension}{Dimension}

  \SetAlgoLined

  $SliceIdStart\leftarrow blockIdx\times SlicePerBlock$\; 
  $SliceIdEnd\leftarrow SliceIdStart + SlicePerBlock$\; 
  $SliceId\leftarrow SliceIdStart+warpIdx$\;
  \nosemic\CachedVec$\leftarrow$\;
  \pushline\nosemic\nonl\InputVector$[partBoundary[blockIdx]$\;
  \dosemic\nonl$\rightarrow partBoundary[blockIdx+1]]$\;
  \popline\While{$SliceId<SliceEndId$}{
    $row\leftarrow warpLane+SliceId\times warpSize$\;
    $Position\leftarrow\PositionELL[SliceId]$\;
    $Width\leftarrow\WidthELL[SliceId]$\;
    \For{$k$ from $1$ to $Width$}{
        $dataIdx\leftarrow Position+warpSize\times k+warpLane$\;
        \nosemic$y+=\ValELL[dataIdx]\times$\;
        \pushline\dosemic\nonl$\CachedVec[colELL[dataIdx]]$\;
    }
    \OutputVector[\row]$\leftarrow y$\;
    \If{$warpLane == 0$}{
     $SliceId\leftarrow$atomicAdd($SliceId$);
    }
    shfl($SliceId$)\;
  }
  $SliceIdER\leftarrow atomicAdd(GlobalSilcedIdER)$\; 
  Repreat Line 5 to 18 on ER part of the matrix\;

  \caption{CUDA kernel function for EHYB framework}
  \label{alg:3}
\end{algorithm}

\section{Evaluation}
In this section, we tested our implementation with 94 large sparse matrices derived from problems related to real-world physical problem simulation in 
various applications. The data used in the experiments were not deliberately selected to fit the algorithms in this paper; rather, we did our best to 
make the matrices used in the experiments come from different domains. The matrices used includes the structural problem, Computational Fluid Dynamic 
problems, electromagnetics problems, biomedical problems, power system, VLSI/semiconductor device problem, semiconductor processes problem, model reduction 
problems, and optimization problem. We exclude the matrices derived from web/DNA connections since the main purpose of this paper is to develop a 
high-performance SpMV framework for iterative solvers of the very large sparse linear system derived from PDE-based problems. For the limitation of space
We list the name of matrices and performance of the EHYB framework on these matrices in the table at the appendix section.   
The experiments in this section are conducted on the SDSC expanse cluster GPU nodes. The server is equipped with NVIDIA V100 SMX2 GPU, 374 GB DDR4 memory 
(running at 2,500 MHz) and a 20 core Intel Xeon Gold 6248 CPU. The GPU nodes we used contains 4 GPUs per node, but all the experiments in this study only
used one GPU for the CUDA kernel function and at most 16 CPU cores for the preprocessing phase of EHYB.\par
To compare with the start-of-the-art performance of the GPU based SpMV, as described in section \ref{section2}, we will compare the EHYB performance with 
the performance of holaspmv, yaspmv (single precision only), CSR5, merge-based SpMV and the latest CUSPARSE generic SpMV interface with ALG1 and ALG2. The 
results is demonstrated in the following sections. 
\subsection{single precision result}
\begin{figure}[ht]
  \centering
  \includegraphics[width=\linewidth]{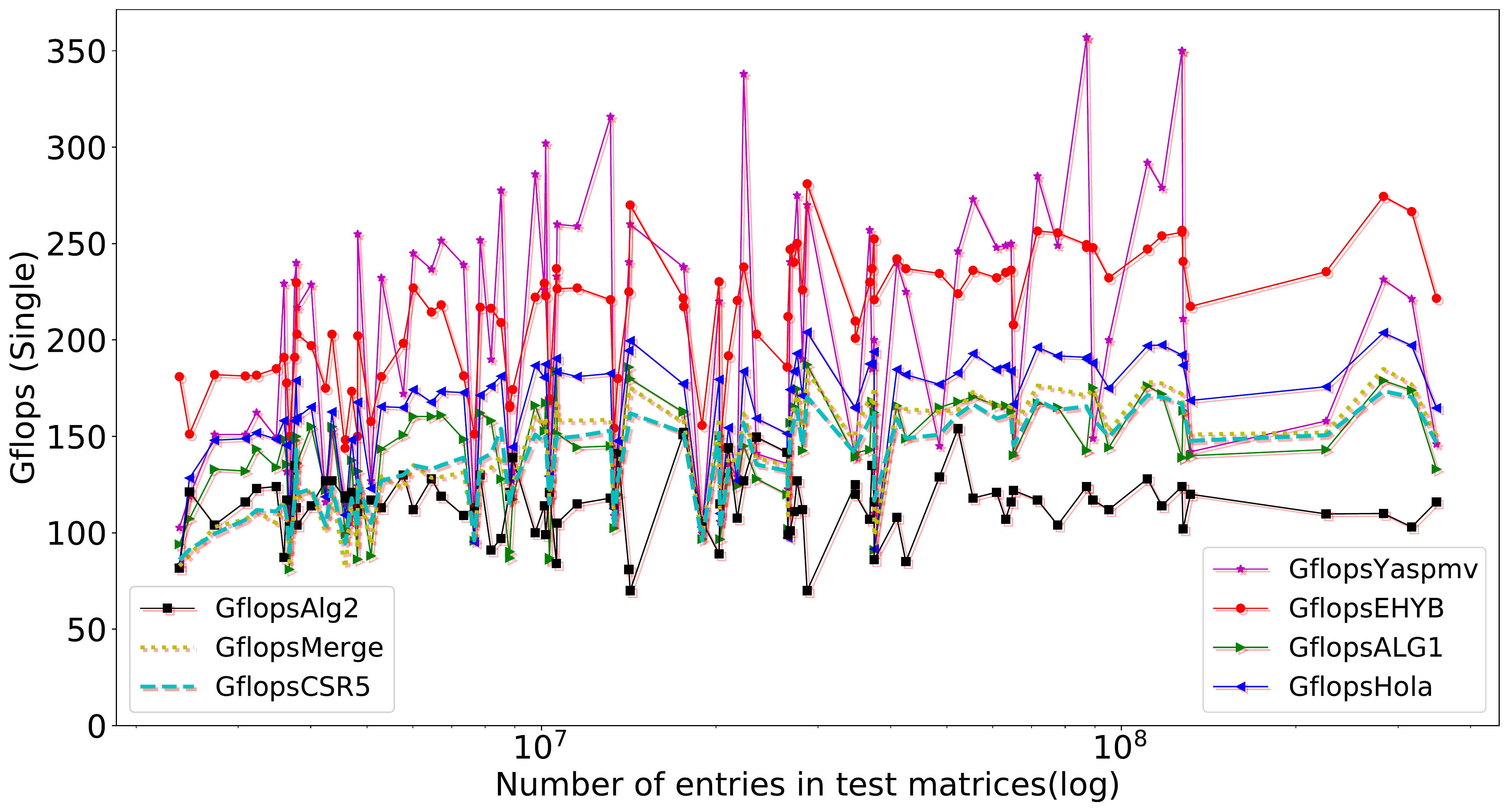}
  \caption{Float precision performance on 94 matrcies from suiteSparse}
  \label{fig:floatLine}
  \Description{Gflop Line graphic for float precision}
\end{figure}

Figure \ref{fig:floatLine} demonstrates the performance of EHYB in single precision, please be aware that the yaspmv didn't generate correct results in 
some of the large matrices (also observed in \cite{steinberger2017globally}), and we changed all \_shfl\_ related function in holaSpMV to the synchronized 
function with the same name because the unsynchronize \_shfl\_ function is no longer supported on CUDA 10.2.
From figure \ref{fig:floatLine} we can find out that EHYB over performs yaSpmv in most test data matrices (60\%), the average speedup of EHYB versus yaSpmv 
is 1.15, and for larger matrices, EHYB performance gain comapred to yaSpmv is more significant. From table \ref{table1} we can also find out holaSpmv is 
the fastest framework which do not requires preprocessing. Averagely EHYB is 1.3$\times$ faster than holaSpmv in single precision.  
\begin{figure}[ht]
  \centering
  \includegraphics[width=\linewidth]{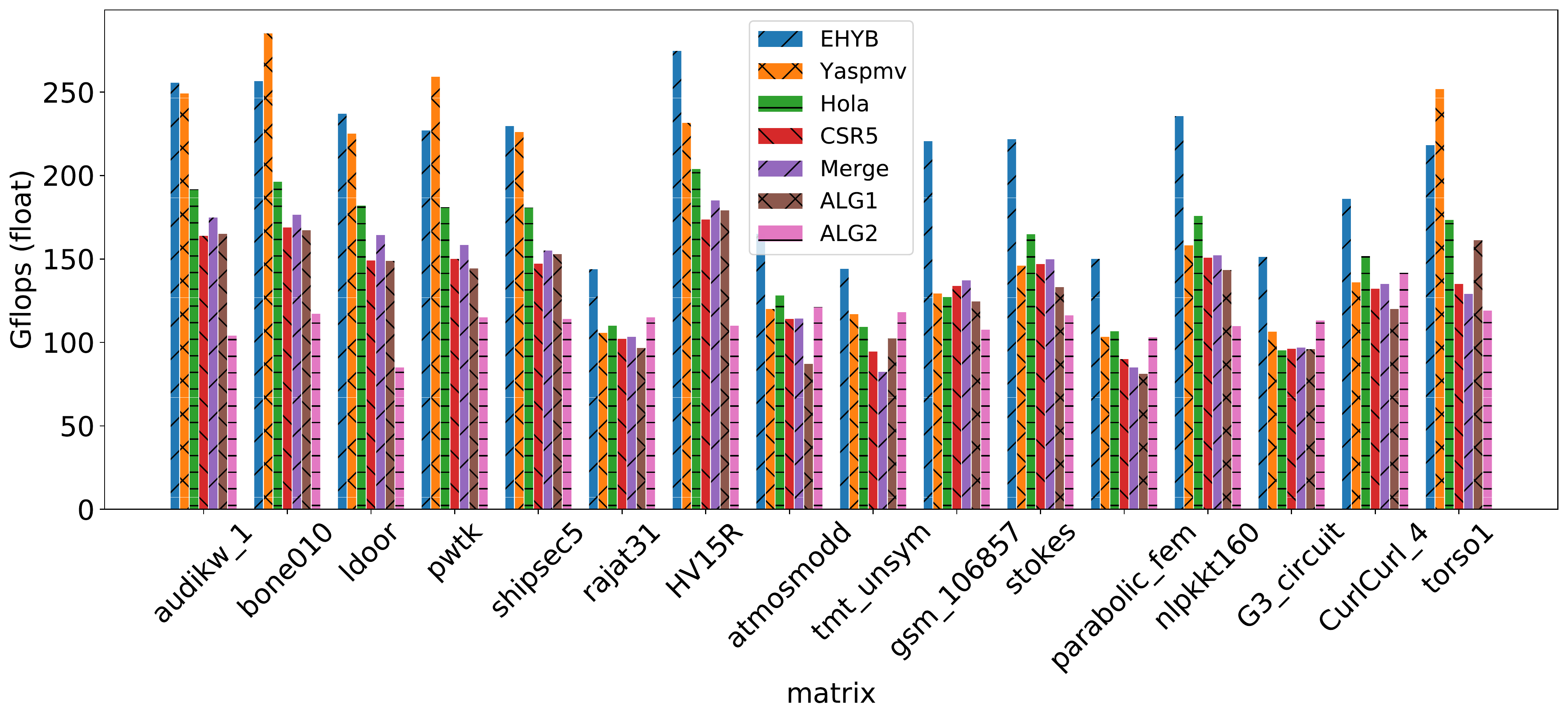}
  \caption{Single precision performance of 16 commonly tested matrices}
  \label{fig:floatBar}
  \Description{float Gflop 16 matrices bar}
\end{figure}

\begin{table*}[ht]
  \begin{center}
    \begin{tabular}{c|c|c|c|c}
      \hline\hline
      \textbf{SpMV framework: } & \textbf{EHYB is faster in \% of matrices} & \textbf{max speedup} & \textbf{min speedup} &\textbf{average speedup}\\
      \hline
       yasmpv & 60.6\% & 2.0 & 0.69 & 1.13\\
      \hline
       holaspmv & 100\% & 2.4 & 1.05& 1.304\\
      \hline
       CSR5 & 100\% &1.89 & 1.305 & 1.53\\
      \hline
       Merge & 100\% & 2.25 & 1.31& 1.517\\
      \hline
       ALG1 & 100\% & 2.42& 1.21&1.518\\
      \hline
       ALG2 & 100\% & 4.0& 1.21& 1.90\\
      \hline
    \end{tabular}
    \caption{Speed up of EHYB versus other frameworks on single precision performance}
    \label{table1}
  \end{center}
\end{table*}

\subsection{double precision result}
\begin{figure}[ht]
  \centering
  \includegraphics[width=\linewidth]{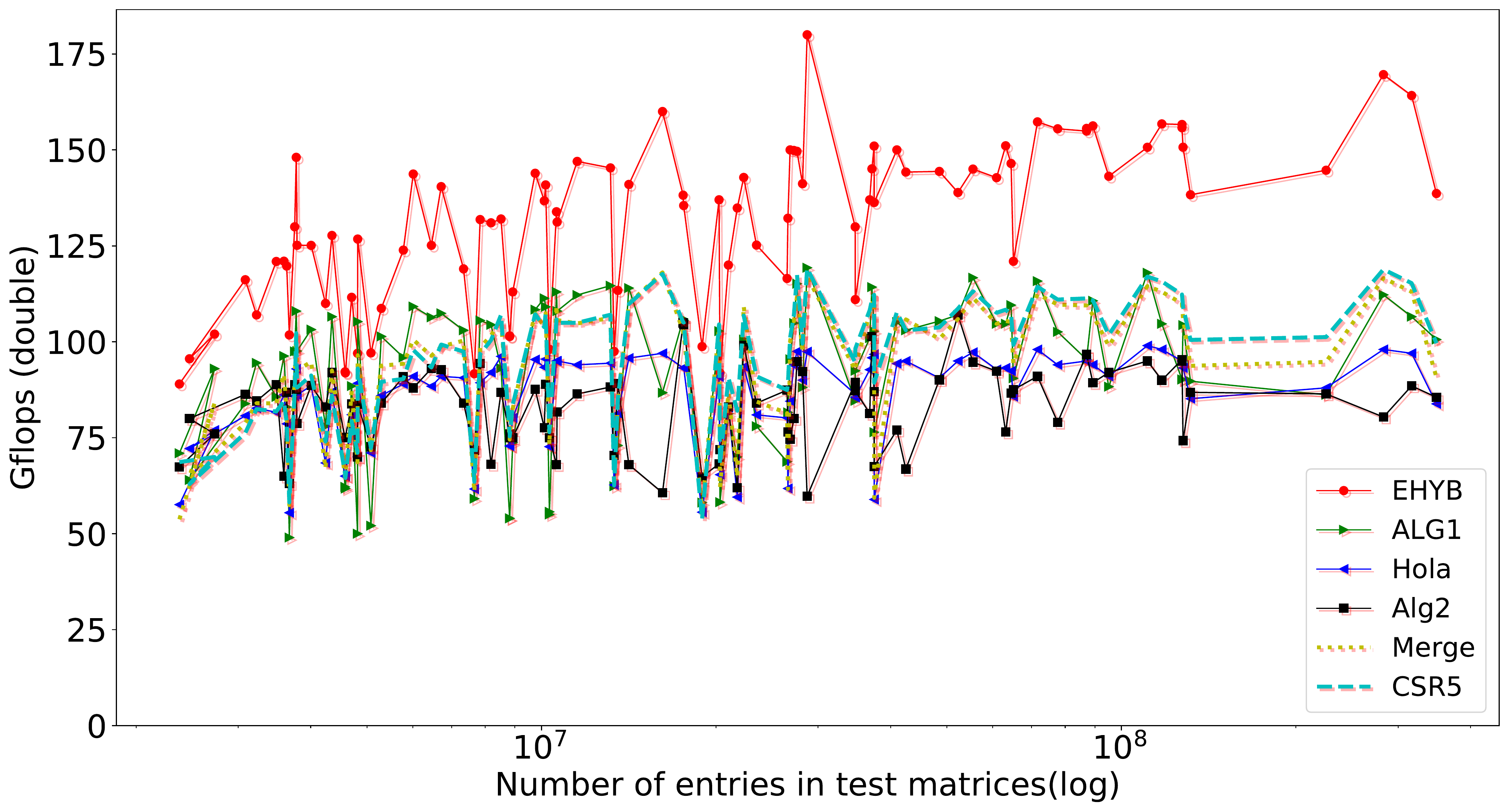}
  \caption{Double precision performance on 94 test matrices}
  \label{fig:doubleLine}
  \Description{Gflop Line graphic for double precision}
\end{figure}

\begin{figure}[ht]
  \centering
  \includegraphics[width=\linewidth]{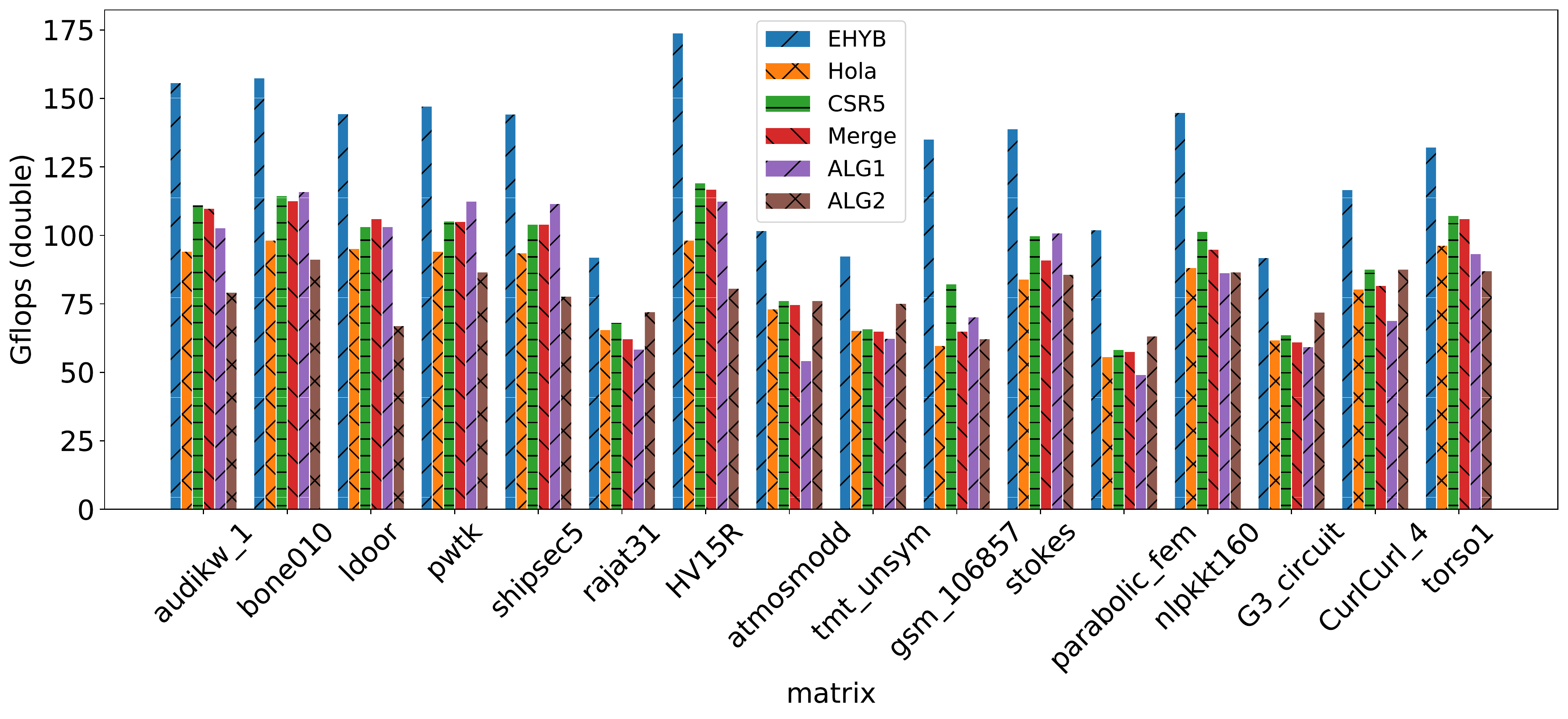}
  \caption{Double precision performance of 16 commonly tested matrices}
  \label{fig:doubleBar}
  \Description{double Gflop 16 matrices bar}
\end{figure}

Figure \ref{fig:doubleLine} and figure \ref{fig:doubleBar} demonstrates the performance of EHYB compared with other implementations. The yaspmv library 
didn't support double precision, so we compared the performance of EHYB with the remaining frameworks.  
From figure \ref{fig:doubleLine} we can find out the EHYB overperforms the other frameworks in all test matrices with a significant performance gain. 
The hola spmv performs solver than fastest CUSPARSE interface at most matrices in double precision tests, which is not match with the reuslts of 
single precision experiment. According to table \ref{table2}, in double precision experiment, the CSR5 framework becomes the fastest framework besides EHYB. 

\begin{table*}[ht]
  \begin{center}
    \begin{tabular}{c|c|c|c}
      \hline\hline
      \textbf{SpMV framework: } &  \textbf{max speedup} & \textbf{min speedup} &\textbf{average speedup}\\
      \hline
       holaspmv &  2.3 & 1.26& 1.5\\
      \hline
       CSR5 & 1.84& 1.13 & 1.38\\
      \hline
       Merge & 2.29 & 1.15& 1.41\\
      \hline
       ALG1 &  2.07& 1.07&1.45\\
      \hline
       ALG2 &  3.01& 1.19& 1.59\\
      \hline
    \end{tabular}
    \caption{Speed up of EHYB versus other frameworks on double precision performance}
    \label{table2}
  \end{center}
\end{table*}
\section{preprocessing time cost and the significance of this study }
\begin{figure}[ht]
  \centering
  \includegraphics[width=\linewidth]{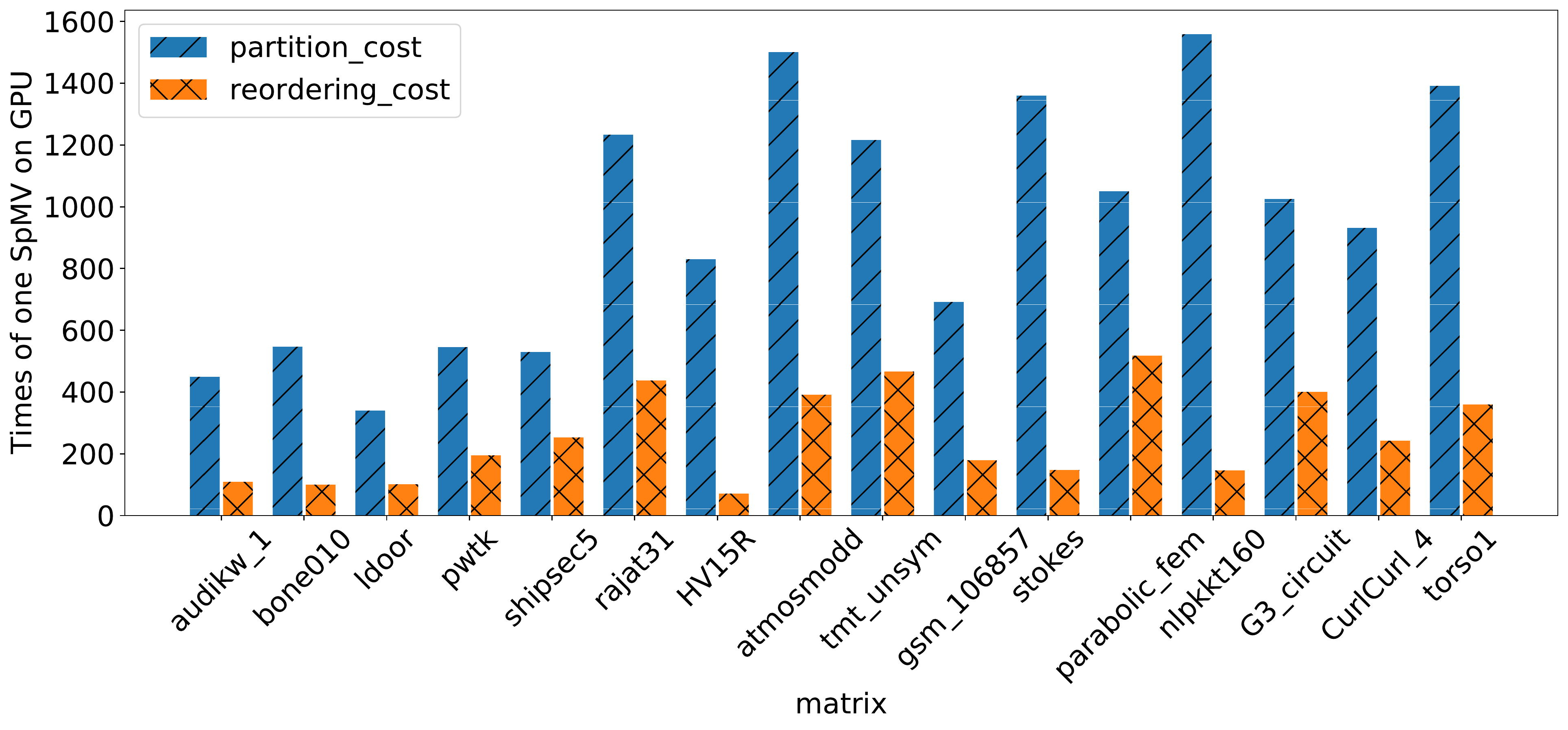}
  \caption{Preprocessing time for 16 commonly tested matrices}
  \label{fig:prepro}
  \Description{preprocessing bar fig}
\end{figure}

Figure \ref{fig:prepro} demonstrates the preprocessing time of EHYB framework for 16 commonly tested sparse matrices. The preprocessing time cost can be 
decomposed into two parts: the partitioning part and the reordering part. In this study, we conducting the METIS partitioning with 16 threads on the host CPU 
(Intel Xeon Gold 6248). From figure \ref{fig:prepro} we can find out that the partitioning time cost for EHYB is around  400$\times$ to 1500$\times$ of 
the time cost of single SpMV operations on the GPU. And the reordering time cost is 50$\times$ to 400$\times$. The total preprocessing time will be around 
500$\times$ to 2000$\times$. We should notice that the main part of the reordering time cost is caused by the sort operations in algorithm \ref{algPrep}, 
currently, we conducting the sort operation with single thread, we expect that the reordering time will reduce significantly if the multi-thread sort 
algorithm is applied.\par
Although EHYB performs significant preprocessing, its preprocessing time cost is still around 100$\times$ less than the yaspmv\cite{yan2014yaspmv} while it
overperforms yaspmv on our experiments. The performance gain is also stable when compared to the time cost machine-learning-based partitiong/reordering
preprocessing technology for GPU based SpMV \cite{dufrechou2021selecting} \cite{benatia2016sparse}. \par 
There are concerns about whether this research will benefit the real-world applications. The major reason for these concerns is: whether the SpMV operations 
still dominate the time cost of solving the linear system in the latest FEM algorithms. Since the modern preconditioned iterative 
solver has a high convergence rate, the solver can find the solution with less iterations, which makes it impossible to amortize the cost of 
preprocessing phase required by EHYB. \par
The widely cited classic literature \cite{saad2003iterative} indicates that the time costs for SpMV dominated the overall time consumed for 
FEM-based simulation for large 3D problems. A recent book\cite{bertaccini2018iterative} also described this property of the FEM applications. On the other hand, in recent 
years, many new methods are developed on building high-efficient preconditioner for large linear systems derived from PDE. The algebraic 
multigrid (AMG) preconditioner\cite{boyle2010hsl_mi20} can significantly improve the convergence rate of the iterative solver, thus reduce the iteration
number to several hundred to several dozens\cite{isotton2021chronos}. When AMG preconditioner is applied, building preconditioner matrices on coarser levels of a 
hierarchy of levels and forming and applying prolongation may cost more time than the the SpMV operations.\par 
However, efficiently parallelize the AMG algorithm with the massive parallel accelerator such as GPGPU is a challange problem\cite{naumov2015amgx}.
In GPGPU platform, the sparse approximate inverse (SPAI) preconditioner works well in many real-world applications according to latest 
literature\cite{doi:10.1177/10943420211017188}\cite{pearson2020preconditioners}\cite{singh2020preconditioned}. The high-performance SpMV kernel proposed in 
this paper will significantly improve the performance of the SPAI preconditioned iterative solver on GPGPU, since the SpMV operation will still be the 
most time consuming part of the iterative solver with SPAI preconditioner\cite{gao2021thread}\cite{georgescu2013gpu}.\par
The SPAI preconditioned iterative solver usually requires thousands of iterations to converge\cite{doi:10.1177/10943420211017188}. Also, in transient simulation, the solver
will repeatly solving the same linear system with hundreds of time steps\cite{kaehler2004transient}. For transient simulation on real-world problems, 
the result of the preprocessing phase in EHYB is shared by hundreds of thousands of iterations. We can safely assume that the preprocessing time cost for 
EHYB can be amortized when applying the SPAI preconditioned solver for transient simulation problems.  
\section{Conclusion}
In this study, we developed a novel SpMY framework on GPU named EHYB. The EHYB framework improves the speed of data movements by partitioning and explicitly 
caching the input vector to the shared memory of the CUDA kernel. The EHYB framework also reduces the volume of data movements by storing the major part of 
the column index in a compact format. The CUDA kernel function in EHYB is optimized to improving the balance between warps of CUDA threads. We tested the
EHYB framework with a large number of sparse matrices derived from real-world applications. The experiment results indicate that the EHYB framework can 
overperform stat-of-the-art SpMV frameworks.  

\begin{acks}
The GPU resources used in this paper is provided by the Extreme Science and Engineering Discovery Environment(XSEDE). 
\end{acks}

\bibliographystyle{ACM-Reference-Format}
\bibliography{sample-base}


\begin{thebibliography}{28}


\ifx \showCODEN    \undefined \def \showCODEN     #1{\unskip}     \fi
\ifx \showDOI      \undefined \def \showDOI       #1{#1}\fi
\ifx \showISBNx    \undefined \def \showISBNx     #1{\unskip}     \fi
\ifx \showISBNxiii \undefined \def \showISBNxiii  #1{\unskip}     \fi
\ifx \showISSN     \undefined \def \showISSN      #1{\unskip}     \fi
\ifx \showLCCN     \undefined \def \showLCCN      #1{\unskip}     \fi
\ifx \shownote     \undefined \def \shownote      #1{#1}          \fi
\ifx \showarticletitle \undefined \def \showarticletitle #1{#1}   \fi
\ifx \showURL      \undefined \def \showURL       {\relax}        \fi
\providecommand\bibfield[2]{#2}
\providecommand\bibinfo[2]{#2}
\providecommand\natexlab[1]{#1}
\providecommand\showeprint[2][]{arXiv:#2}

\bibitem[\protect\citeauthoryear{Anzt, Cojean, Yen-Chen, Dongarra, Flegar,
  Nayak, Tomov, Tsai, and Wang}{Anzt et~al\mbox{.}}{2020a}]%
        {anzt2020load}
\bibfield{author}{\bibinfo{person}{Hartwig Anzt}, \bibinfo{person}{Terry
  Cojean}, \bibinfo{person}{Chen Yen-Chen}, \bibinfo{person}{Jack Dongarra},
  \bibinfo{person}{Goran Flegar}, \bibinfo{person}{Pratik Nayak},
  \bibinfo{person}{Stanimire Tomov}, \bibinfo{person}{Yuhsiang~M Tsai}, {and}
  \bibinfo{person}{Weichung Wang}.} \bibinfo{year}{2020}\natexlab{a}.
\newblock \showarticletitle{Load-balancing sparse matrix vector product kernels
  on GPUs}.
\newblock \bibinfo{journal}{\emph{ACM Transactions on Parallel Computing
  (TOPC)}} \bibinfo{volume}{7}, \bibinfo{number}{1} (\bibinfo{year}{2020}),
  \bibinfo{pages}{1--26}.
\newblock


\bibitem[\protect\citeauthoryear{Anzt, Tomov, and Dongarra}{Anzt
  et~al\mbox{.}}{2014}]%
        {anzt2014implementing}
\bibfield{author}{\bibinfo{person}{Hartwig Anzt}, \bibinfo{person}{Stanimire
  Tomov}, {and} \bibinfo{person}{Jack Dongarra}.}
  \bibinfo{year}{2014}\natexlab{}.
\newblock \showarticletitle{Implementing a Sparse Matrix Vector Product for the
  SELL-C/SELL-C-$\sigma$ formats on NVIDIA GPUs}.
\newblock \bibinfo{journal}{\emph{University of Tennessee, Tech. Rep.
  ut-eecs-14-727}} (\bibinfo{year}{2014}).
\newblock


\bibitem[\protect\citeauthoryear{Anzt, Tsai, Abdelfattah, Cojean, and
  Dongarra}{Anzt et~al\mbox{.}}{2020b}]%
        {anzt2020evaluating}
\bibfield{author}{\bibinfo{person}{Hartwig Anzt}, \bibinfo{person}{Yuhsiang~M
  Tsai}, \bibinfo{person}{Ahmad Abdelfattah}, \bibinfo{person}{Terry Cojean},
  {and} \bibinfo{person}{Jack Dongarra}.} \bibinfo{year}{2020}\natexlab{b}.
\newblock \showarticletitle{Evaluating the Performance of NVIDIA’s A100
  Ampere GPU for Sparse and Batched Computations}. In
  \bibinfo{booktitle}{\emph{2020 IEEE/ACM Performance Modeling, Benchmarking
  and Simulation of High Performance Computer Systems (PMBS)}}. IEEE,
  \bibinfo{pages}{26--38}.
\newblock


\bibitem[\protect\citeauthoryear{Bell and Garland}{Bell and Garland}{2009}]%
        {bell2009implementing}
\bibfield{author}{\bibinfo{person}{Nathan Bell} {and} \bibinfo{person}{Michael
  Garland}.} \bibinfo{year}{2009}\natexlab{}.
\newblock \showarticletitle{Implementing sparse matrix-vector multiplication on
  throughput-oriented processors}. In \bibinfo{booktitle}{\emph{Proceedings of
  the conference on high performance computing networking, storage and
  analysis}}. \bibinfo{pages}{1--11}.
\newblock


\bibitem[\protect\citeauthoryear{Benatia, Ji, Wang, and Shi}{Benatia
  et~al\mbox{.}}{2016}]%
        {benatia2016sparse}
\bibfield{author}{\bibinfo{person}{Akrem Benatia}, \bibinfo{person}{Weixing
  Ji}, \bibinfo{person}{Yizhuo Wang}, {and} \bibinfo{person}{Feng Shi}.}
  \bibinfo{year}{2016}\natexlab{}.
\newblock \showarticletitle{Sparse matrix format selection with multiclass SVM
  for SpMV on GPU}. In \bibinfo{booktitle}{\emph{2016 45th International
  Conference on Parallel Processing (ICPP)}}. IEEE, \bibinfo{pages}{496--505}.
\newblock


\bibitem[\protect\citeauthoryear{Bertaccini and Durastante}{Bertaccini and
  Durastante}{2018}]%
        {bertaccini2018iterative}
\bibfield{author}{\bibinfo{person}{Daniele Bertaccini} {and}
  \bibinfo{person}{Fabio Durastante}.} \bibinfo{year}{2018}\natexlab{}.
\newblock \bibinfo{booktitle}{\emph{Iterative methods and preconditioning for
  large and sparse linear systems with applications}}.
\newblock \bibinfo{publisher}{CRC Press}.
\newblock


\bibitem[\protect\citeauthoryear{Boyle, Mihajlovi{\'c}, and Scott}{Boyle
  et~al\mbox{.}}{2010}]%
        {boyle2010hsl_mi20}
\bibfield{author}{\bibinfo{person}{Jonathan Boyle}, \bibinfo{person}{Milan
  Mihajlovi{\'c}}, {and} \bibinfo{person}{Jennifer Scott}.}
  \bibinfo{year}{2010}\natexlab{}.
\newblock \showarticletitle{HSL\_MI20: an efficient AMG preconditioner for
  finite element problems in 3D}.
\newblock \bibinfo{journal}{\emph{International journal for numerical methods
  in engineering}} \bibinfo{volume}{82}, \bibinfo{number}{1}
  (\bibinfo{year}{2010}), \bibinfo{pages}{64--98}.
\newblock


\bibitem[\protect\citeauthoryear{Cevahir, Nukada, and Matsuoka}{Cevahir
  et~al\mbox{.}}{2009}]%
        {cevahir2009fast}
\bibfield{author}{\bibinfo{person}{Ali Cevahir}, \bibinfo{person}{Akira
  Nukada}, {and} \bibinfo{person}{Satoshi Matsuoka}.}
  \bibinfo{year}{2009}\natexlab{}.
\newblock \showarticletitle{Fast conjugate gradients with multiple GPUs}. In
  \bibinfo{booktitle}{\emph{International Conference on Computational
  Science}}. Springer, \bibinfo{pages}{893--903}.
\newblock


\bibitem[\protect\citeauthoryear{Dufrechou, Ezzatti, and
  Quintana-Ort{\'\i}}{Dufrechou et~al\mbox{.}}{2021}]%
        {dufrechou2021selecting}
\bibfield{author}{\bibinfo{person}{Ernesto Dufrechou}, \bibinfo{person}{Pablo
  Ezzatti}, {and} \bibinfo{person}{Enrique~S Quintana-Ort{\'\i}}.}
  \bibinfo{year}{2021}\natexlab{}.
\newblock \showarticletitle{Selecting optimal SpMV realizations for GPUs via
  machine learning}.
\newblock \bibinfo{journal}{\emph{The International Journal of High Performance
  Computing Applications}} (\bibinfo{year}{2021}),
  \bibinfo{pages}{1094342021990738}.
\newblock


\bibitem[\protect\citeauthoryear{Gao, Chen, and He}{Gao et~al\mbox{.}}{2021}]%
        {gao2021thread}
\bibfield{author}{\bibinfo{person}{Jiaquan Gao}, \bibinfo{person}{Qi Chen},
  {and} \bibinfo{person}{Guixia He}.} \bibinfo{year}{2021}\natexlab{}.
\newblock \showarticletitle{A thread-adaptive sparse approximate inverse
  preconditioning algorithm on multi-GPUs}.
\newblock \bibinfo{journal}{\emph{Parallel Comput.}}  \bibinfo{volume}{101}
  (\bibinfo{year}{2021}), \bibinfo{pages}{102724}.
\newblock


\bibitem[\protect\citeauthoryear{Georgescu, Chow, and Okuda}{Georgescu
  et~al\mbox{.}}{2013}]%
        {georgescu2013gpu}
\bibfield{author}{\bibinfo{person}{Serban Georgescu}, \bibinfo{person}{Peter
  Chow}, {and} \bibinfo{person}{Hiroshi Okuda}.}
  \bibinfo{year}{2013}\natexlab{}.
\newblock \showarticletitle{GPU acceleration for FEM-based structural
  analysis}.
\newblock \bibinfo{journal}{\emph{Archives of Computational Methods in
  Engineering}} \bibinfo{volume}{20}, \bibinfo{number}{2}
  (\bibinfo{year}{2013}), \bibinfo{pages}{111--121}.
\newblock


\bibitem[\protect\citeauthoryear{Isotton, Frigo, Spiezia, and Janna}{Isotton
  et~al\mbox{.}}{2021a}]%
        {isotton2021chronos}
\bibfield{author}{\bibinfo{person}{Giovanni Isotton}, \bibinfo{person}{Matteo
  Frigo}, \bibinfo{person}{Nicol{\`o} Spiezia}, {and} \bibinfo{person}{Carlo
  Janna}.} \bibinfo{year}{2021}\natexlab{a}.
\newblock \showarticletitle{Chronos: A general purpose classical AMG solver for
  High Performance Computing}.
\newblock \bibinfo{journal}{\emph{arXiv preprint arXiv:2102.07417}}
  (\bibinfo{year}{2021}).
\newblock


\bibitem[\protect\citeauthoryear{Isotton, Janna, and Bernaschi}{Isotton
  et~al\mbox{.}}{2021b}]%
        {doi:10.1177/10943420211017188}
\bibfield{author}{\bibinfo{person}{Giovanni Isotton}, \bibinfo{person}{Carlo
  Janna}, {and} \bibinfo{person}{Massimo Bernaschi}.}
  \bibinfo{year}{2021}\natexlab{b}.
\newblock \showarticletitle{A GPU-accelerated adaptive FSAI preconditioner for
  massively parallel simulations}.
\newblock \bibinfo{journal}{\emph{The International Journal of High Performance
  Computing Applications}} \bibinfo{volume}{0}, \bibinfo{number}{0}
  (\bibinfo{year}{2021}), \bibinfo{pages}{01--12}.
\newblock
\urldef\tempurl%
\url{https://doi.org/10.1177/10943420211017188}
\showDOI{\tempurl}
\showeprint{https://doi.org/10.1177/10943420211017188}


\bibitem[\protect\citeauthoryear{Kaehler and Henneberger}{Kaehler and
  Henneberger}{2004}]%
        {kaehler2004transient}
\bibfield{author}{\bibinfo{person}{Christian Kaehler} {and}
  \bibinfo{person}{Gerhard Henneberger}.} \bibinfo{year}{2004}\natexlab{}.
\newblock \showarticletitle{Transient 3-D FEM computation of eddy-current
  losses in the rotor of a claw-pole alternator}.
\newblock \bibinfo{journal}{\emph{IEEE Transactions on magnetics}}
  \bibinfo{volume}{40}, \bibinfo{number}{2} (\bibinfo{year}{2004}),
  \bibinfo{pages}{1362--1365}.
\newblock


\bibitem[\protect\citeauthoryear{Li, Yang, and Li}{Li et~al\mbox{.}}{2014}]%
        {li2014performance}
\bibfield{author}{\bibinfo{person}{Kenli Li}, \bibinfo{person}{Wangdong Yang},
  {and} \bibinfo{person}{Keqin Li}.} \bibinfo{year}{2014}\natexlab{}.
\newblock \showarticletitle{Performance analysis and optimization for SpMV on
  GPU using probabilistic modeling}.
\newblock \bibinfo{journal}{\emph{IEEE Transactions on Parallel and Distributed
  Systems}} \bibinfo{volume}{26}, \bibinfo{number}{1} (\bibinfo{year}{2014}),
  \bibinfo{pages}{196--205}.
\newblock


\bibitem[\protect\citeauthoryear{Liu and Vinter}{Liu and Vinter}{2015}]%
        {liu2015csr5}
\bibfield{author}{\bibinfo{person}{Weifeng Liu} {and} \bibinfo{person}{Brian
  Vinter}.} \bibinfo{year}{2015}\natexlab{}.
\newblock \showarticletitle{CSR5: An efficient storage format for
  cross-platform sparse matrix-vector multiplication}. In
  \bibinfo{booktitle}{\emph{Proceedings of the 29th ACM on International
  Conference on Supercomputing}}. \bibinfo{pages}{339--350}.
\newblock


\bibitem[\protect\citeauthoryear{Merrill and Garland}{Merrill and
  Garland}{2016}]%
        {merrill2016merge}
\bibfield{author}{\bibinfo{person}{Duane Merrill} {and}
  \bibinfo{person}{Michael Garland}.} \bibinfo{year}{2016}\natexlab{}.
\newblock \showarticletitle{Merge-based sparse matrix-vector multiplication
  (SpMV) using the CSR storage format}.
\newblock \bibinfo{journal}{\emph{ACM SIGPLAN Notices}} \bibinfo{volume}{51},
  \bibinfo{number}{8} (\bibinfo{year}{2016}), \bibinfo{pages}{1--2}.
\newblock


\bibitem[\protect\citeauthoryear{Mittal and Vetter}{Mittal and Vetter}{2015}]%
        {mittal2015survey}
\bibfield{author}{\bibinfo{person}{Sparsh Mittal} {and}
  \bibinfo{person}{Jeffrey~S Vetter}.} \bibinfo{year}{2015}\natexlab{}.
\newblock \showarticletitle{A survey of CPU-GPU heterogeneous computing
  techniques}.
\newblock \bibinfo{journal}{\emph{ACM Computing Surveys (CSUR)}}
  \bibinfo{volume}{47}, \bibinfo{number}{4} (\bibinfo{year}{2015}),
  \bibinfo{pages}{1--35}.
\newblock


\bibitem[\protect\citeauthoryear{Naumov, Arsaev, Castonguay, Cohen, Demouth,
  Eaton, Layton, Markovskiy, Reguly, Sakharnykh, et~al\mbox{.}}{Naumov
  et~al\mbox{.}}{2015}]%
        {naumov2015amgx}
\bibfield{author}{\bibinfo{person}{Maxim Naumov}, \bibinfo{person}{M Arsaev},
  \bibinfo{person}{Patrice Castonguay}, \bibinfo{person}{J Cohen},
  \bibinfo{person}{Julien Demouth}, \bibinfo{person}{Joe Eaton},
  \bibinfo{person}{S Layton}, \bibinfo{person}{N Markovskiy},
  \bibinfo{person}{Istv{\'a}n Reguly}, \bibinfo{person}{Nikolai Sakharnykh},
  {et~al\mbox{.}}} \bibinfo{year}{2015}\natexlab{}.
\newblock \showarticletitle{AmgX: A library for GPU accelerated algebraic
  multigrid and preconditioned iterative methods}.
\newblock \bibinfo{journal}{\emph{SIAM Journal on Scientific Computing}}
  \bibinfo{volume}{37}, \bibinfo{number}{5} (\bibinfo{year}{2015}),
  \bibinfo{pages}{S602--S626}.
\newblock


\bibitem[\protect\citeauthoryear{Nisa, Siegel, Rajam, Vishnu, and
  Sadayappan}{Nisa et~al\mbox{.}}{2018}]%
        {nisa2018effective}
\bibfield{author}{\bibinfo{person}{Israt Nisa}, \bibinfo{person}{Charles
  Siegel}, \bibinfo{person}{Aravind~Sukumaran Rajam}, \bibinfo{person}{Abhinav
  Vishnu}, {and} \bibinfo{person}{P Sadayappan}.}
  \bibinfo{year}{2018}\natexlab{}.
\newblock \showarticletitle{Effective machine learning based format selection
  and performance modeling for SpMV on GPUs}. In \bibinfo{booktitle}{\emph{2018
  IEEE International Parallel and Distributed Processing Symposium Workshops
  (IPDPSW)}}. IEEE, \bibinfo{pages}{1056--1065}.
\newblock


\bibitem[\protect\citeauthoryear{Pearson and Pestana}{Pearson and
  Pestana}{2020}]%
        {pearson2020preconditioners}
\bibfield{author}{\bibinfo{person}{John~W Pearson} {and}
  \bibinfo{person}{Jennifer Pestana}.} \bibinfo{year}{2020}\natexlab{}.
\newblock \showarticletitle{Preconditioners for Krylov subspace methods: An
  overview}.
\newblock \bibinfo{journal}{\emph{GAMM-Mitteilungen}} \bibinfo{volume}{43},
  \bibinfo{number}{4} (\bibinfo{year}{2020}), \bibinfo{pages}{e202000015}.
\newblock


\bibitem[\protect\citeauthoryear{Saad}{Saad}{2003}]%
        {saad2003iterative}
\bibfield{author}{\bibinfo{person}{Yousef Saad}.}
  \bibinfo{year}{2003}\natexlab{}.
\newblock \bibinfo{booktitle}{\emph{Iterative methods for sparse linear
  systems}}.
\newblock \bibinfo{publisher}{SIAM}.
\newblock


\bibitem[\protect\citeauthoryear{Singh and Ahuja}{Singh and Ahuja}{2020}]%
        {singh2020preconditioned}
\bibfield{author}{\bibinfo{person}{Navneet~Pratap Singh} {and}
  \bibinfo{person}{Kapil Ahuja}.} \bibinfo{year}{2020}\natexlab{}.
\newblock \showarticletitle{Preconditioned linear solves for parametric model
  order reduction}.
\newblock \bibinfo{journal}{\emph{International Journal of Computer
  Mathematics}} \bibinfo{volume}{97}, \bibinfo{number}{7}
  (\bibinfo{year}{2020}), \bibinfo{pages}{1484--1502}.
\newblock


\bibitem[\protect\citeauthoryear{Steinberger, Zayer, and Seidel}{Steinberger
  et~al\mbox{.}}{2017}]%
        {steinberger2017globally}
\bibfield{author}{\bibinfo{person}{Markus Steinberger}, \bibinfo{person}{Rhaleb
  Zayer}, {and} \bibinfo{person}{Hans-Peter Seidel}.}
  \bibinfo{year}{2017}\natexlab{}.
\newblock \showarticletitle{Globally homogeneous, locally adaptive sparse
  matrix-vector multiplication on the GPU}. In
  \bibinfo{booktitle}{\emph{Proceedings of the International Conference on
  Supercomputing}}. \bibinfo{pages}{1--11}.
\newblock


\bibitem[\protect\citeauthoryear{Yan, Li, Zhang, and Zhou}{Yan
  et~al\mbox{.}}{2014}]%
        {yan2014yaspmv}
\bibfield{author}{\bibinfo{person}{Shengen Yan}, \bibinfo{person}{Chao Li},
  \bibinfo{person}{Yunquan Zhang}, {and} \bibinfo{person}{Huiyang Zhou}.}
  \bibinfo{year}{2014}\natexlab{}.
\newblock \showarticletitle{yaSpMV: Yet another SpMV framework on GPUs}.
\newblock \bibinfo{journal}{\emph{Acm Sigplan Notices}} \bibinfo{volume}{49},
  \bibinfo{number}{8} (\bibinfo{year}{2014}), \bibinfo{pages}{107--118}.
\newblock


\bibitem[\protect\citeauthoryear{Yang, Li, and Li}{Yang et~al\mbox{.}}{2017}]%
        {yang2017hybrid}
\bibfield{author}{\bibinfo{person}{Wangdong Yang}, \bibinfo{person}{Kenli Li},
  {and} \bibinfo{person}{Keqin Li}.} \bibinfo{year}{2017}\natexlab{}.
\newblock \showarticletitle{A hybrid computing method of SpMV on CPU--GPU
  heterogeneous computing systems}.
\newblock \bibinfo{journal}{\emph{J. Parallel and Distrib. Comput.}}
  \bibinfo{volume}{104} (\bibinfo{year}{2017}), \bibinfo{pages}{49--60}.
\newblock


\bibitem[\protect\citeauthoryear{Yang, Li, and Li}{Yang et~al\mbox{.}}{2018}]%
        {yang2018parallel}
\bibfield{author}{\bibinfo{person}{Wangdong Yang}, \bibinfo{person}{Kenli Li},
  {and} \bibinfo{person}{Keqin Li}.} \bibinfo{year}{2018}\natexlab{}.
\newblock \showarticletitle{A parallel computing method using blocked format
  with optimal partitioning for SpMV on GPU}.
\newblock \bibinfo{journal}{\emph{J. Comput. System Sci.}}
  \bibinfo{volume}{92} (\bibinfo{year}{2018}), \bibinfo{pages}{152--170}.
\newblock


\bibitem[\protect\citeauthoryear{Zheng, Gu, Gu, Yang, and Liu}{Zheng
  et~al\mbox{.}}{2014}]%
        {zheng2014biell}
\bibfield{author}{\bibinfo{person}{Cong Zheng}, \bibinfo{person}{Shuo Gu},
  \bibinfo{person}{Tong-Xiang Gu}, \bibinfo{person}{Bing Yang}, {and}
  \bibinfo{person}{Xing-Ping Liu}.} \bibinfo{year}{2014}\natexlab{}.
\newblock \showarticletitle{BiELL: A bisection ELLPACK-based storage format for
  optimizing SpMV on GPUs}.
\newblock \bibinfo{journal}{\emph{J. Parallel and Distrib. Comput.}}
  \bibinfo{volume}{74}, \bibinfo{number}{7} (\bibinfo{year}{2014}),
  \bibinfo{pages}{2639--2647}.
\newblock


\end{thebibliography}

\appendix
\section{Source code}
The source code of this paper is available at:
\url{https://github.com/Chong-Chen-UNLV/EHYB_SPMV_GPU}
\section{Matrices used for testing}

\begin{table*}[h]
    \fontsize{8}{10}\selectfont
    \begin{tabular}{c|c|c|c|c|c|c|c}
      \hline\hline
      \textbf{Matrix name} &  \textbf{Category} & \textbf{Dimension} &\textbf{Entries}&\textbf{Matrix name} &  \textbf{Category} & \textbf{Dimension} &\textbf{Entries}\\
      \hline
        poisson3D & CFD & 85,623 & 2,374,949 &ship\_003	& Structural & 121,728 &8,086,034 \\
      \hline
        atmosmodj & CFD & 1,270,432  & 8,814,880 &BenElechi1	& 3D Problem &245,874 &13,150,496 \\
      \hline
        vas\_stokes\_1M & VLSI &1,090,664 &34,767,207&Hook\_1498  & Structural& 1,498,023&60,917,445\\
      \hline
        CurlCurl\_1 &Model Reduction &226,451 &2,472,071&laminar\_duct3D& CFD& 67,173& 3,833,077\\
      \hline
        CurlCurl2& Model Reduction& 806,529&8,921,789&memchip &Circuit Simulation &2,707,524&14,810,202 \\
      \hline
        inline1 & Structural &503,712 &36,816,342 &Geo\_1438 &Structural &1,437,960&63,156,690 \\
      \hline
        windtunnel\_evap3d 	&CFD &40,816 &2,730,600&cant&3D problem &62,451&4,007,383 \\
      \hline
        m\_t1	& Structural & 97,578&9,753,570&CurlCurl\_3	 &Model Reduction &1,219,574&13,544,618 \\
      \hline
        PFlow\_742 & 3D problem&742,793 &37,138,461 &Serena  &Structural  &1,391,349&64,131,971 \\
      \hline
        cfd2	&CFD & 123,440&3,087,898&offshore & Electromagnetics &259,789&4,242,673 \\
      \hline
        shipsec5	&Structural  & 179,860&10,113,096&crankseg\_2 & structural &63,838&14,148,858 \\
      \hline
        RM07 & CFD &381,689 &37,464,962&vas\_stokes\_2M  &Semiconductor  &2,146,677&65,129,037 \\
      \hline
      Goodwin\_095&CFD & 100,037&3,226,066&t3dh	 &Model Reduction &79,171&4,352,105 \\
      \hline
      x104	& Structural & 108,384&10,167,624&TSOPF\_RS\_b2383\_c1&power net &38,120&16,171,169 \\
      \hline
       nv2  &Semiconductor &1,453,908 &52,728,362&bone010 & Bio Engineering &986,703&71,666,325 \\
      \hline
        FEM\_3D\_thermal2	&Thermal &147,900 &3,489,300&af\_4\_k101 &Structural &503,625& 17,550,675\\
      \hline
       atmosmodl  &CFD &1,489,752 &10,319,760&audikw\_1  &Structural  &943,695&77,651,847 \\
      \hline
       Emilia\_923 &Structural  &923,136 & 41,005,206&t2em &Electromagnetics  &921,632&4,590,832 \\
      \hline
       oilpan &Structural  & 73,752& 3,597,188&af\_shell8\_9\_10 &Structural  &1,508,065&52,672,325 \\
      \hline
%
         atmosmodm	&CFD &1,489,752 & 10,319,760&consph	 &3D problem & 83,334& 6,010,480\\
      \hline
         ldoor& Structural &952,203 & 46,522,475&Transport &Structural  &1,602,111 &23,500,731 \\
      \hline
        Dubcova3&3D Problem &146,689 & 3,636,649&Cube\_Coup\_dt6  & Structural & 2,164,760& 127,206,144\\
      \hline
         crankseg\_1	&Structural  &52,804 & 10,614,210&TEM152078	 &Electromagnetics  & 152,078& 6,459,326\\
      \hline
        dielFilterV2real  &Electromagnetics &1,157,456 &48,538,952& CurlCurl\_4&Model Reduction &806,529 &8,921,789 \\
      \hline
        parabolic\_fem	&CFD &525,825 & 3,674,625&Bump\_2911 &3D problem &2,911,419 &127,729,899 \\
      \hline
         bmwcra\_1	& Structural & 148,770&10,641,602 &boneS01 &bio Engineering & 127,224& 6,715,152\\
      \hline
        tmt\_unsym & Electromagnetics &917,825 &4,584,801 &dgreen	 & Semiconductor &1,200,611 & 38,259,877\\
      \hline
         s3dkt3m2	&Structural  & 90,449&4,820,891 &vas\_stokes\_4M  & Semiconductor &4,382,246
 & 131,577,616\\
      \hline
        pwtk	& Structural &217,918 &11,634,424 &bmw7st\_1&Structural  & 141,347&7,339,667 \\
      \hline
        boneS10  &Bio Engineering & 914,898& 55,468,422&F1	&Structural  &343,791 & 26,837,113\\
      \hline
        Long\_Coup\_dt0 & Structural &1,470,152 & 87,088,992& nlpkkt160 &Optimization  & 8,345,600& 229,518,112\\
      \hline
         engine	&Structural  &143,571 &4,706,073 &G3\_circuit	 &Circuit Simulation & 1,585,478& 7,660,826\\
      \hline
         Freescale1	& Circuit Simulation&3,428,755 & 18,920,347&Fault\_639	 &Structural  &638,802 &28,614,564 \\
      \hline
         Long\_Coup\_dt6 &Structural  &638,802 &28,614,564 &HV15R  &CFD &2,017,169 &283,073,458 \\
      \hline
         apache2	&Structural  & 715,176& 4,817,870&TEM181302	 &Electromagnetics  & 181,302&7,839,010 \\
      \hline
         msdoor	& Structural &415,863 & 19,173,163&ML\_Laplace	&Structural  &377,002 &27,689,972 \\
      \hline
         dielFilterV3real &Electromagnetics  &1,102,824 & 89,306,020& Queen\_4147 &3D Problem & 4,147,110
& 329,499,284 \\
      \hline
        s3dkq4m2	& Structural & 90,449 & 4,820,891&PR02R	 &CFD &161,070 & 8,185,136\\
      \hline
        rajat31	& Circuit Simulation& 4,690,002&20,316,253 &nlpkkt80 &Optimization  &1,062,400 &28,704,672 \\
      \hline
        nlpkkt120	&Optimization  & 3,542,400& 96,845,792& stokes &Semiconductor  &11,449,533 & 349,321,980\\
      \hline
        StocF-1465	& CFD & 1,465,137&21,005,389 & torso1	&Bio Engineering &116,158 &8,516,500 \\
      \hline
        ML\_Geer & Structural &1,504,002 &110,879,972 &	tmt\_sym& Electromagnetics &726,713 & 5,080,961\\
      \hline
      F2	& Structural &71,505 & 5,294,285&atmosmodd	&CFD & 1,270,432&8,814,880 \\
      \hline
       gsm\_106857	& Electromagnetics&589,446 &21,758,924 & ss &Semiconductor  &1,652,680 &34,753,577 \\
      \hline
      Flan\_1565 & Structural & 1,564,794&117,406,044 & Cube\_Coup\_dt0 &Structural  &2,164,760 & 124,406,070\\
      \hline
       Goodwin\_127	& Structural & 178,437&5,778,545 & CoupCons3D	&Structural  &416,800 &22,322,336 \\
      \hline
      \end{tabular}
    \caption{Matrix tested in this paper }
    \label{tableAppendix1}
\end{table*}

\end{document}